\magnification=\magstep1
\advance\voffset by 3truemm
\advance\hoffset by -1truemm
\vsize=23truecm
\hsize=16.5truecm
\overfullrule=0pt
\hfuzz 5truept
\parskip=5pt
\baselineskip=12pt
\font\abstractfont=cmr9
\font\abstracttitlefont=cmti9
\font\authorfont=cmr10 scaled\magstep1
\font\sectionfont=cmbx10 scaled\magstep2
\font\titlefont=cmbx10 scaled\magstep2
\def\title#1{\null\vskip26truemm\noindent{\titlefont#1}}
\def\author#1{\vskip15truemm\noindent{\authorfont#1}}
\def\address#1{\vskip5truemm\noindent#1}
\def\abstract#1{\vskip26truemm\noindent{\abstracttitlefont Abstract.}
{\abstractfont#1}\vskip16truemm}
\newcount\numsection
\def\references{\bigskip\bigbreak\noindent\leftline{\bf REFERENCES}\bigskip}
\def\section#1{\advance\numsection by1
\bigskip\bigbreak\noindent\leftline{\sectionfont\the\numsection\
#1}\bigskip\nobreak}
%
%
\let\epsilon=\varepsilon
\def\d{{\rm d}}
\def\e{{\rm e}}
\def\i{{\rm i}}
\def\frac#1#2{\hbox{$#1\over#2$}}
\def\mat#1{\left[\matrix{#1}\right]}
\leftline{Preprint UGVA-DPT 05-927/96}
\title{Quantum Theory without Quantification\footnote{$\ast$}{{\rm
For the Hepp-Hunziker volume of Helvetica Physica Acta}}
}
\author{C. Piron}
\address{D\'epartement de Physique Th\'eorique, 24 quai Ernest-Ansermet,
CH-1211 Gen\`eve 4}
\abstract{After having explained Samuel Clarke's conception of the new
philosophy of physical reality, we will treat the electron field in this
context as a field modifying the void. From this we will be able to
derive the so-called quantum rules just from Noether's theorem on
conserved currents. Thus quantum theory appears as a kind of nonlocal
field theory, in fact a new theory.}
\section{Introduction}
In spite of much remarkable progress, the fashionable physicist's window
on world reality has not changed during the past century. It remains
always the same, the philosophical view of Descartes and Leibnitz: the
reality of the universe is nothing else than a ``nothing'', the void --
just a recipient filled with little particles and other kinds of ether
[Le Sage 1818]. In
spite of the official claim that obviously fields and quantum particles
have replaced such primitive and outmoded concepts, in fact most physicists
continue to think in the same terms. They imagine gravitation and
electromagnetism as deformations of some substantial ether with a lot
of vibrations and corresponding waves. Quantum phenomena reduce to a
manifestation of stochastic motions and path averages of little
particles [Piron 1995].
To get out of this rut and refuse to allow paradoxes such as infinities and
inconsistencies which bog down theoretical physics, we must absolutely
forsake this inadequate received view and accept as fundamental
the views of Newton and Clarke. First of all, we must admit the separate
existence of space and time. As Clarke says [1866], the void space is not an
attribute without subject but a space without bodies. In other words the
void exists in of itself: after all it has properties, it has place, it
has three dimensions, and it is Euclidean (or almost Euclidean). At
first sight, it may
seem that here we run into the difficulty of how to verify such
properties, since if we introduce some apparatus then we no longer have
the void. Such apparent difficulties have been solved by Dirk Aerts by
his formulation of the notion of element of reality and of how the
notion of definite experimental project gives a criterion to check the
existence of such an element [Aerts 1982]. By experimental project, Aerts
means an experiment that one can very well eventually perform on the system,
and where the positive result has been defined in advance. Following
Einstein and Aerts, we will say that the system has an element of
reality, or an actual property, if in the event of the corresponding
experiment the positive result would be certainly obtained. As we can
see, an element of reality is an actual property of the system itself
which exists even in the absence of any apparatus.
The void in a vessel has a volume of one litre if one could exactly fill
it with one litre of water if one decided to check its volume. Of course
the water-filled vessel is not a vessel filled with the void. When we
claim that the void here is Euclidean, we claim the existence of an
element of reality, since if one were to construct a triangle here with
three solid rods then certainly the sum of the angles would be found to
be 180$^\circ\,$. The void space itself has such a Euclidean property
in the absence of any rod. As a third example, we claim that ``the void
space at this moment has here an electric field'' means that if one
decided to place here an electric charge, the charge would be certainly
be accelerated. Such a field is an element of reality of the void space
in the absence of the test charge: when one makes the experiment to
verify the existence of the field, of course one completely destroys the
situation and in fact that field in the presence of the charge is even
not defined.
Changing one's mind and accepting the existence of void space for itself
is not enough however, one must also accept time and the flow of time
also as having a separate existence in themselves. By nature, space and
time are completely different, each possible place in space is actual in
this moment, but for time only the present is actual -- the future can
be partially chosen and will become actual, whereas the past which has been
cannot be
changed in any manner whatsoever. The arrow of time exists. This can be
checked with the following simple experiment. Let a ball bounce up and down
and predict in the middle whether it will arrive at the bottom or the top
(as you can see this is a symmetric situation). One can affirm arrival
at the top or the bottom after the fact but not before, since in the latter
case you could always stop the motion of the ball with your hand. This
translates causality and exhibits the arrow of time.
To recapitulate, to describe the world as it really is we must introduce,
from the very beginning, both the void space and the time, and so
consider particles or other objects as manifestations of space. In this
spirit, a particle is considered not as a manifestation of some
substance existing alone and by itself, but as some manifest property of
the void space intrinsically connected with its
surroundings. In this context, the description of a particle in
the classical approximation cannot be given just by the position
of some hypothetical object, but must be at least completed by specifying the
state of the surroundings, that is, the momentum. This is the physical
explanation of the 7-dimensional space in classical mechanics.
A charge (say an electron) is here in the void space
surrounded by its field (the electric displacement $\vec{D}\,(x)$ in the
electron case). Such a field reacts with the void generating another
field, a field of force (the electric field $\vec{E}\,(x)\,$) which can act
on other particles. As we see, even in the vacuum such an action is
never direct: it is a major physical error to identify $\vec{E}\,(x)$
and $\vec{D}\,(x)$ even in the vacuum.
The fundamental fact that some objects called particles are entities (that
is cannot be subdivided) and are not localised at a point but occupy place,
in other words are not local (a very reasonable hypothesis in our window), is
the fundamental fact which explains and justifies the rules of quantum
physics. In the following we will present a model of the electron field as an
example to illustrate this new philosophy. We will take from the very
beginning the point of view explained just above with the goal of justifying
the quantum rules just from such a notion of field.
\section{Construction of coordinates for space and time}
Before introducing the notion of particle, we must give a mathematical
model for the void space and the time themselves. We have already
treated such a problem in Helvetica Physica Acta [Barut {\it et al.\/} 1993].
Here we will not repeat
such a construction, but will explain in more detail its physical origin.
We build spatial coordinates from a chosen inertial reference frame, which is
nothing else than a solid object sufficiently large to not be perturbed by
small fluctuations of its surroundings: for example, one could take the
``earth'' freely falling onto the sun (just perturbed a little by the moon),
but a hypothetical earth which does not rotate. Having in this way constructed
${\bf R}^3\,$, a model for space, we add a fourth coordinate, the time $t\,$,
given by a clock at rest in the chosen reference frame, thereby obtaining
${\bf R}^4\,$, a model for space and time. According to Einstein's principle
of relativity, any such ${\bf R}^4$ is as good as any other ${\bf R}'{}^4$
obtained using another inertial reference frame. The coordinates in
${\bf R}^4$ and ${\bf R}'{}^4$ are related by the usual Lorentz
transformation, however to interpret such a transformation we must come back
from ${\bf R}^4$ and ${\bf R}'{}^4$ to the sole physical space and the
physical time with its sole actual instant [Piron 1990].
\section{The electron as a field}
As we have said in the introduction, fields and particles manifest
properties of the existing void during the flow of time. Having built a
model ${\bf R}^4$ for space and time, physical fields and particles will be
described by mathematical fields on ${\bf R}^4\,$, for instance spinor fields
$\phi(x)$ or covector fields $A_\mu(x)\,$. In our example, the electron
field will be of such kind, in fact a four component complex spinor field
$\Psi(x)$ (and $\Psi^\dagger(x)\,$). Such a field is of course in interaction
with other fields such as the electromagnetic and gravitational fields
among others which, for simplicity, will be considered as given independently
of the electron field $\Psi$ itself. For this very reason we will call them
exterior fields. As we have developed elsewhere [Piron and Moore 1995], an
adequate formalism here is the Cartan formalism, where one introduces a 4-form
on a mathematical space $\Sigma\,$, here ${\bf R}^{12}\,$, which must reflect
all possible states of our electron field model.
More precisely, the space $\Sigma$ is built with the four coordinates $x^\mu$
of a space and time model, together with the four complex components (eight
real numbers) of the column matrix $\Psi$ which describes the electron field.
The equations for the electron field $\Psi$ (sometimes called improperly
equations of ``motion'' or equations of ``propagation'') are determined by
the Cartan principle, which affirms that
$s^\ast(i_{\scriptscriptstyle X}\,d\omega)=0$
for such a field $s:{\bf R}^4\rightarrow\Sigma\,$, where $X$ runs over all
tangent vector fields on $\Sigma\,$. As we will see, the model is then
completely determined by the chosen 4-form $\omega\,$:
$$\omega=-eA_\mu(x)\Psi^\dagger\alpha^\mu\Psi\,\eta
-\i\hbar\Psi^\dagger\alpha^\mu\,d\Psi\wedge\eta_\mu
+m\Psi^\dagger\beta\Psi\,\eta\,$$
where $A_\mu(x)$ is the (given) electromagnetic field, and for notational
convenience we have introduced the usual odd forms
$$\eqalign{
\eta&=\frac{1}{24}\,\epsilon_{\mu\nu\rho\lambda}\,dx^\mu\wedge dx^\nu\wedge
dx^\rho\wedge dx^\lambda\,,\cr
\eta_\mu&=\frac{1}{6}\,\epsilon_{\mu\nu\rho\lambda}\,dx^\nu\wedge dx^\rho\wedge
dx^\lambda\,,\cr
\eta_{\mu\nu}&=\frac{1}{2}\,\epsilon_{\mu\nu\rho\lambda}\,dx^\rho\wedge
dx^\lambda\,.\cr
}$$
The $\alpha^\mu$ and $\beta$ are five $4\!\times\!4$ matrices which express the
dynamical covariance of the theory. In the Lorentz case, according to Dirac
we choose
$$\alpha^0=\mat{I&0\cr0&I}\,,\qquad
\alpha^i=c\mat{0&\sigma^i\cr\sigma^i&0\cr}\,,\qquad
\beta=c^2\mat{I&0\cr0&-I\cr}\,,$$
but in the Galilean approximation (the usual Schr\"odinger case)
according to Levy-Leblond we will choose
$$\alpha^0=\mat{I&0\cr0&0}\,,\qquad
\alpha^i=\mat{0&\sigma^i\cr\sigma^i&0\cr}\,,\qquad
\beta=2\mat{0&0\cr0&-I}\,.$$
As we see, the 4-form $\omega$ contains three terms, each one with its own
constant prefactor (the electric charge $e\,$, the Planck constant
$\hbar\,$, the electron mass $m\,$) which is clearly a little redundant. Again,
we insist that all fields are fields of properties of the existing void.
Let us first derive the equation for the field $\Psi\,$. We have
$$d\omega=-eA_\mu(x)\,d\big(\Psi^\dagger\alpha^\mu\Psi\big)\wedge\eta
-\i\hbar\,d\Psi^\dagger\wedge\alpha^\mu\,d\Psi\wedge\eta_\mu
+m\,d(\Psi^\dagger\beta\Psi)\wedge\eta\,.$$
Choosing the tangent vector field $X=\hat{e}{}_{\Psi^\dagger}$ we obtain
$$\eqalign{
s^\ast\big(i_{\hat{e}{}_{\Psi^\dagger}}d\omega\big)&=
s^\ast\big(-eA_\mu(x)\alpha^\mu\Psi\,\eta
-\i\hbar\alpha^\mu\,d\Psi\wedge\eta_\mu+m\beta\Psi\,\eta\big)\cr
&=\big(-eA_\mu(x)\alpha^\mu\Psi(x)-\i\hbar\alpha^\mu\partial_\mu\Psi(x)
+m\beta\Psi(x)\big)s^\ast(\eta)\cr
&=0\,.\cr
}$$
Since $s^\ast(\eta)\not=0$ we then have that
$$\Big[\alpha^\mu\big(-\i\hbar\partial_\mu-eA_\mu(x)\big)+
m\beta\Big]\Psi(x)=0\,.$$
In the Lorentz case this is exactly the Dirac equation. In the Galilean
approximation, we find upon rewriting
$\Psi=\mat{\phi\cr\chi\cr}$
$$\eqalign{
\i\hbar\partial_0\phi(x)&=\sigma^j\big(-\i\hbar\partial_j-eA_j(x)\big)\chi(x)
-eA_0(x)\phi(x)\,,\cr
0&=\sigma^j\big(-\i\hbar\partial_j-eA_j(x)\big)\phi(x)-2m\chi(x)\,.\cr
}$$
Substituting for $\chi(x)$ we then have
$$\i\hbar\partial_0\phi(x)=\Big[\frac{1}{2m}g^{jk}
\big(-\i\hbar\partial_j-eA_j(x)\big)\big(-\i\hbar\partial_k-eA_k(x)\big)
+\frac{\hbar e}{2m}\sigma_iB^i(x)-eA_0(x)\Big]\phi(x)\,,$$
which is exactly the two component (Pauli-)Schr\"odinger equation with the
good factor $\frac{\hbar e}{2m}$ coupling the spin $\sigma^i$ to the magnetic
field $B^i(x)=\partial_jA_k(x)-\partial_kA_j(x)\,$.
\section{Noether's theorem and the quantum interpretation}
One success of the Cartan formalism is the derivation of the Noether
theorem it affords. Consider a one parameter group acting on the state
space $\Sigma\,$. Suppose that this group is generated by a vector field
on $\Sigma\,$:
$$g_\lambda^\ast:r\mapsto r+\lambda Y\,,$$
where here $r$ is a point in $\Sigma\,$, not to be confused with the $x$
in ${\bf R}^4\,$, and $g_\lambda$ is the action of the germ of
the group. Now suppose that the 4-form $\omega$ is invariant under the
action of $g_\lambda$ so that $g_\lambda^\ast\omega=\omega\,$. Then
as is well known
$$L_{\scriptscriptstyle Y}\omega=i_{\scriptscriptstyle Y}d\omega+
di_{\scriptscriptstyle Y}\omega=0\,,$$
where $L_{\scriptscriptstyle Y}$ is the Lie derivative by $Y\,$.
The Noether theorem affirms that the 3-form $i_{\scriptscriptstyle Y}\omega$
is conserved on the solution $s\,$:
$$ds^\ast(i_{\scriptscriptstyle Y}\omega)=0\,.$$
The proof is straightforward, since
$$ds^\ast(i_{\scriptscriptstyle Y}\omega)
=s^\ast(di_{\scriptscriptstyle Y}\omega)
=s^\ast(-i_{\scriptscriptstyle Y}d\omega)=0\,.$$
The first equality is trivial, the second translates the invariance of
$\omega\,$, and the third is just the field equation.
To interpret such a result mathematically, we write the 3-form
$s^\ast(i_{\scriptscriptstyle Y}\omega)$ as
$$s^\ast(i_{\scriptscriptstyle Y}\omega)=J^\mu(x)\,\eta_\mu\,.$$
It is then easy to recognise that the current $J^\mu(x)$ is conserved, since
the Noether theorem gives
$$\partial_\mu J^\mu(x)=0\,.$$
To remove one difficulty of the conventional (outmoded) field theory, we
want to emphasise that we have derived the conserved currents from the
Noether theorem without requiring some physical symmetry of the field
equations, using merely some formal invariance of our 4-form. This is
particularly transparent in the next example, where only a gauge
invariance of the {\it first\/} kind is invoked. Indeed, it is easy to check
that our $\omega$ is invariant under the transformations
$$\Psi^\dagger\mapsto\Psi^\dagger\e^{-\i\lambda}\qquad\hbox{and}\qquad
\Psi\mapsto\e^{\i\lambda}\Psi\,,$$
which are generated by the tangent vector field
$$Y=-\i\Psi^\dagger\hat{e}{}_{\Psi^\dagger}+\i\hat{e}{}_\Psi\Psi\,.$$
The Noether theorem then gives the following conserved current
$$s^\ast(i_{\scriptscriptstyle Y}\omega)=\hbar\,\Psi^\dagger(x)\alpha^\mu
\Psi(x)\,\eta_\mu\,.$$
In the Lorentz case, this is nothing else than the conservation of the
scalar product
$$\textstyle{\int\limits_{\scriptscriptstyle{\bf R}^3}}\Psi^\dagger(x)\alpha^0
\Psi(x)\,\eta_0=\textstyle{\int\limits_{\scriptscriptstyle{\bf R}^3}}
\Psi^\dagger(x)\Psi(x)\,\d V\,,$$
which is exactly the scalar product used by Dirac to solve the one
body hydrogen atom problem. We must remark that with the usual argument this
supposes that the integral exists and that $\psi(x)$ tends to $0$ at infinity.
On the other hand, since the norm is invariant the linear field equation
can be proved to induce a unitary transformation via the so called Wigner
theorem.
In the case of the Galilean approximation, the corresponding
conserved scalar product is
$$\textstyle{\int\limits_{\scriptscriptstyle{\bf R}^3}}\Psi^\dagger(x)\alpha^0
\Psi(x)\,\eta_0=\textstyle{\int\limits_{\scriptscriptstyle{\bf R}^3}}
\phi^\dagger(x)\phi(x)\,\d V\,.$$
Here only the two first components play a role, exactly justifying the
``prescription of quantum mechanics'' and the use of the ``two component
Schr\"odinger wave equation for spin $\frac12\,$''.
As the reader can remark, to obtain quantum theory we have
utilised neither ``quantum prescriptions'' nor the ``correspondance
principle'', but just Noether's theorem applied to field in
vacuum space (sic). We can also justify in the same way other ``quantum
rules'', for instance the momentum operator. In the free case ($A_\mu(x)=0\,$),
which we will consider for the rest of this section, our 4-form $\omega$ is
invariant by the action of the passive space translations
$$x^j\mapsto x^j+\lambda h^j\,,$$
where the $h^j$ are three numbers, the generators of the translation in
the inertial reference frame defining ${\bf R}^3\,$. If the translation
is just in the direction $\hat{e}{}_j\,$, one of the unit vectors of
${\bf R}^3\,$, we can simply write
$$x^j\mapsto x^j+\lambda\,,$$
and the generator in the state $\Sigma$ is just the vector field
$\hat{e}{}_j\,$. In such a
case, the conserved current given by Noether's theorem is
$$\eqalign{
s^\ast(i_{\hat{e}{}_j}\omega)&=s^\ast\big(\i\hbar\Psi^\dagger\alpha^\mu
d\Psi\wedge\eta_{\mu j}+m\Psi^\dagger\beta\Psi\,\eta_j\big)\cr
&=\i\hbar\Psi^\dagger(x)\alpha^\mu\partial_j\Psi(x)\,\eta_\mu
-\i\hbar\Psi^\dagger(x)\alpha^\mu\partial_\mu\Psi(x)\,\eta_j
+m\Psi^\dagger(x)\beta\Psi(x)\,\eta_j\,,\cr
}$$
which, taking into account the electron field equation, gives simply
$$s^\ast(i_{\hat{e}{}_j}\omega)=
\i\hbar\Psi^\dagger(x)\alpha^\mu\partial_j\Psi(x)\,\eta_\mu\,.$$
With the same hypothesis as for the scalar product, this means that
$$\textstyle{\int\limits_{\scriptscriptstyle{\bf R}^3}}\Psi^\dagger(x)\alpha^0
(-\i\hbar\partial_j)\Psi(x)\,\eta_0$$
is conserved. Physically this is the total momentum of the field, and the
corresponding density is given by the well known momentum operator
$$p_j=-\i\hbar\partial_j\,.$$
Our 4-form is also invariant invariant by the action of the passive time
translations, leading in the same way to conservation of the total energy
of the field
$$\textstyle{\int\limits_{\scriptscriptstyle{\bf R}^3}}\Psi^\dagger(x)\alpha^0
(\i\hbar\partial_t)\Psi(x)\,\eta_0
=\textstyle{\int\limits_{\scriptscriptstyle{\bf R}^3}}\Psi^\dagger(x)\big[
\alpha^i(-\i\hbar\partial_i)+m\beta\big]\Psi(x)\,\eta_0\,.$$
As a final example, consider the passive rotations $\lambda$ about the
$\hat{e}{}_i$ axis. Our 4-form is invariant by such
transformations, which by definition act simultaneously on $x\,$, $\Psi$
et $\Psi^\dagger\,$. The corresponding generator is the vector field
$$Y=x_j\hat{e}{}_k-x_k\hat{e}{}_j-\i\hat{e}{}_{\Psi}\,\frac12\,\sigma^i\Psi
+\i\Psi^\dagger\,\frac12\,\sigma^i\hat{e}{}_{\Psi^\dagger}$$
since $\psi$ and $\psi^\dagger$ are spinors. The Noether theorem gives
the conservation of the total angular momentum
$$\textstyle{\int\limits_{{\bf R}^3}}\big[
\Psi^\dagger(x)\alpha^0x_j(-\i\hbar\partial_k)\Psi(x)
-\Psi^\dagger(x)\alpha^0x_k(-\i\hbar\partial_j)\Psi(x)
+\Psi^\dagger(x)\frac{\hbar}{2}\,\sigma^i\Psi(x)\big]\,\eta_0$$
and the corresponding density is the angular momentum operator
$$x_jp_k-x_kp_j+\frac{\hbar}{2}\,\sigma^i\,.$$
As we can see, this operator decomposes into an orbital part
$$x_jp_k-x_kp_j$$
and a highly non-local part, the spin
$$\frac{\hbar}{2}\,\sigma^i\,.$$
\section{Conclusion}
If one accepts this new window, the existence of the void space and one
of its manifestations, the four component complex spinor field $\Psi\,$,
one is led to the fundamental rules of ``quantum mechanics''. These turn
out to be nothing else than the rules of a non-local physics. With
such a $\Psi(x)$ we have built a Hilbert space whose rays are the states
of the field, the electron field. Our position operator $x^j$ and our
momentum operator $p_j$ are the usual ones satisfying the Heisenberg
commutation relations. It is our deep conviction that the particle aspect
of the theory arises only from interactions and selection rules due to
the conservation laws, above all the angular momentum, which contains
a highly non-local term, the spin.
Finally, I would like to express my thanks to David Moore for his
help and encouragement in the difficult process of translating my
thoughts into English.
\references
\parindent=0pt
\def\refjl#1#2#3#4#5{\hangafter=1\hangindent=30pt
#1; ``#2'' {\it #3} {\bf #4} #5.\par}
\def\refbk#1#2#3{\hangafter=1\hangindent=30pt
#1; ``#2'' #3.\par}

\refjl{D. Aerts 1982}{Description of many separated physical entities without
the paradoxes encountered in quantum mechanics}{Found. Phys.}{12}{1131-1170}
\refjl{A. O. Barut, D. J. Moore and C. Piron 1993}{The Cartan formalism in
field theory}{Helv. Phys. Acta}{66}{795-809}
\refbk{S. Clarke 1866 in}{Oeuvres philosophiques de Leibnitz
t.$\,$II}{Quatri\`eme replique de Clarke \`a Leibnitz, \S9, p.643.
Librarie Philosophique de Ladrange (Paris)}
\refbk{Le Sage 1818}{Trait\'e de m\'ecanique physique}{Pr\'evost (Gen\`eve)}
\refbk{C. Piron 1990}{M\'ecanique quantique bases et applications}{Chap. 6.
Presses polytechniques et universitaires romandes (Lausanne)}
\refjl{C. Piron and D. J. Moore 1995}{New aspects of field theory}{Turk. J.
Phys.}{19}{202-216}
\bye